# Transient Magnetoelastic Coupling in CrSBr

Youn Jue Bae,[1,*,†] Taketo Handa,[1] Yanan Dai,[1] Jue Wang,[1] Huicong Liu,[2] Allen Scheie,[3] Daniel G. Chica,[1] Michael E. Ziebel,[1] Andrew D. Kent,[4] Xiaodong Xu,[5] Ka Shen,[2] Xavier Roy,[1] Xiaoyang Zhu[1,*]

[1] Department of Chemistry, Columbia University, New York, NY 10027, USA
[2] The Center of Advanced Quantum Studies and Department of Physics, Beijing Normal University, Beijing 100875, China
[3] MPA-Q, Los Alamos National Laboratory, Los Alamos, NM 87545, USA
[4] Department of Physics, New York University, New York, NY 10003, USA
[5] Department of Physics, University of Washington, Seattle, WA 98195, USA

**ABSTRACT**

**Recent research has revealed remarkable properties of the two-dimensional (2D) van der Waals layered crystal CrSBr, which is both a semiconductor and an A-type antiferromagnet. Here we show the role of strong magnetoelastic coupling in the generation and propagation of coherent magnons in CrSBr. Time and spatially resolved magneto-optical Kerr effect (tr-MOKE) microscopy reveals two time-varying transient strain fields induced by out-of-plane transverse and in-plane longitudinal lattice displacements. These transient strain fields launch coherent wavepackets of magnons, optical and acoustic at 24.6 ± 0.8 GHz and 33.4 ± 0.5 GHz, respectively. These findings suggest mechanisms for controlling and manipulating coherent magnons from distinct magnetoelastic couplings in this 2D van der Waals magnetic semiconductor.**

[*] To whom correspondence should be addressed. Emails: YB: yb293@cornell.edu; XYZ: xyzhu@columbia.edu
[†] Current address: Department of Chemistry and Chemical Biology, Cornell University, Ithaca, NY 14853, USA

# 1. INTRODUCTION

Coherent magnons, also called spin waves, are intrinsically non-linear due to magnetic dipole coupling [1] and their dispersions are tunable by the magnitude and directions of an external magnetic field [2]. These properties make coherent magnons ideal candidates for wave-based logic, directional coupling, and computing [3]. There are various ways to excite coherent magnons, e.g., by microwave antennas, electrical pulses, and optical pulses [4]. Among these different approaches, optical excitation provides easy access without the need for device fabrication and, when combined with optical detection, allows ultrafast tracking of spin dynamics in both time and space domains [5,6]. Optical excitation and detection also resolve low wavevector windows that are not easily accessible with neutron scattering [7]. While optical excitation can directly transfer angular momentum to the spins via the inverse Faraday effect or the Cotton-Mouton effect, a more common mechanism is to exert transient torques on the spin system from a gradient in demagnetization field created by the excitation laser beam profile [8,9]. Alternatively, magnetoelastic coupling can also launch coherent magnons [1].

In this work, we address the mechanisms in the launch and propagation of coherent magnons in CrSBr, a two-dimensional (2D) van der Waals (vdW) material which is both a semiconductor and an A-type antiferromagnet (AFM), i.e., with intralayer ferromagnetic and interlayer antiferromagnetic order [10,11]. The A-type AFM order persists from the bulk vdW crystal to the 2D bilayer limit, with little change in the Néel temperature (132-140 K) [11]. The electronic structure is found to be strongly coupled to magnetic order in CrSBr [12], permitting the detection of coherent magnons in the microwave region by excitonic transitions in the visible to near-infrared region [13]. Such optical access to coherent magnons has been utilized in the tuning of magnon-exciton interactions by external magnetic field [14] and in the imaging of propagating magnons [15]. In the latter work, Sun et al. combined optical imaging on CrSBr with theory and concluded that long range magnetic dipole-dipole coupling dominates the fast propagation of coherent magnons [15]. Here we focus on the launching mechanisms for coherent magnons in CrSBr. Using time-resolved magneto-optical Kerr effect (tr-MOKE) imaging, we identify two time-varying transient strain fields induced by out-of-plane transverse and in-plane longitudinal lattice displacements. In the time-domain experimental data, we observe two distinct time scales in spin wave decay dynamics and attribute these to two different launching mechanisms, namely the laser-induced demagnetization field and the transient strain fields that stem from the intrinsically strong magnetoelastic coupling in CrSBr.

## 2. EXPERIMENTAL

The synthesis, exfoliation, and characterization of CrSBr crystals have been detailed elsewhere [10,11]. Atomically thin flakes of CrSBr can be prepared by mechanical exfoliation onto Si/SiO$_2$ substrates, where the bulk magnetic structure is maintained down to the ferromagnetic (FM) monolayer with a Curie temperature $T_C$ = 146 K and to the antiferromagnetic (AFM) bilayer with a Néel temperature $T_N$ = 140 K [16]. CrSBr is also a direct-gap semiconductor down to the monolayer, with an electronic gap of 1.5 eV and an excitonic gap of 1.34 eV [17]. The anisotropic nature of the exfoliated crystal allows the easy determination of crystallographic axis, *c*- normal to the surface (*z*), *b*- in plane short axis (*x*), *a*- in plane long axis (*y*) [11]. The magnetic structure of CrSBr is depicted in Fig. 1a where the easy, intermediate, and hard axes are along the *b*-, *a*- and *c*- axis, respectively. To perform tr-MOKE imaging (Fig. 1a-1b), we split the output of a Ti:sapphire regenerative amplifier (Coherent RegA, 250 kHz, 800 nm, 100 fs pulse width) into two beams: one for the 740 nm pump generated in an OPA and the other for the 880 nm probe from white-light generation in a sapphire crystal with bandpass filter of 70 nm width. The sample is mounted in a closed-cycle cryostat (Montana Instruments) to maintain a temperature of $T \sim 5$ K. We apply a magnetic field of $\mu_0 H = 0.2$ T along the *c*-axis using a permanent magnet to tilt the spins to the out-of-plane direction necessary for the polar MOKE geometry (Fig. 1a). The pump-probe pulses are delayed up to 7 ns. The probe beam is spatially separated from the pump beam by a dual-axis galvo mirror scanning system for imaging with diffraction-limited resolution (Fig. 1b) [18]. The pump and the probe polarizations are along the *b*-axis of the crystal. A balanced detection scheme is used with lock-in amplifier where the reference signal is synchronized with the optical chopper frequency. In MOKE and transient reflectance, we detect the angle of rotation ($\Delta\phi$) and the magnitude of reflectance ($\Delta R/R$, where R is reflectance without pump and $\Delta R$ is change in reflectance by the pump).

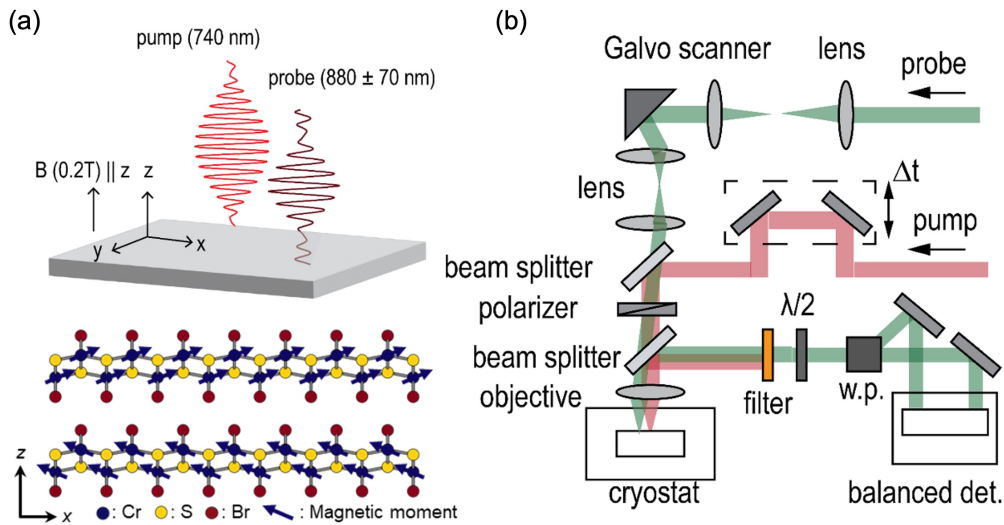

**Fig. 1. Experimental setup for MOKE imaging**. (a) Crystal structure of CrSBr and experimental geometry. The blue arrows represent the magnetic moments which are slightly canted away from the easy *b*-axis in the external magnetic field. (b) Optical set up of the tr-MOKE imaging system. In all experiments, the sample temperature is 5 K and the external field is 0.2T along z.

## 3. RESULTS AND DISCUSSIONS

We optically excite a ~150 nm thick flake of CrSBr by a pump pulse with above gap photon energy $h\nu_1 = 1.67$ eV (pulse width, 150 fs; power 1 µW; Gaussian width 1.4 µm) and collect the spatial map of probe polarization rotation, i.e., polar MOKE imaging, by scanning the linearly polarized probe ($h\nu_2 = 1.4$ eV, pulse width 100 fs; power 0.2 µW; Gaussian width d ~ 1.2 µm) as a function of pump-probe delay ($\Delta t$). The angle of polarization rotation ($\Delta\phi$) is proportional to the pump-induced changes in the out-of-plane magnetization along the *z*-axis ($\Delta m_z$) [19]. Two representative MOKE images at pump-probe time delays of $\Delta t = 20$ ps and 100 ps are shown in Fig. 2a and 2b, respectively. The MOKE images reveal distinct symmetries at short and long delay times. At $\Delta t = 20$ ps, the image presents a dipolar pattern in $\Delta m_z$; it is symmetric and antisymmetric with respect to the *xz* and *xy* mirror planes, respectively. At the longer delay of $\Delta t = 100$ ps, the image evolves to a quadrupolar pattern which is antisymmetric with respect to both *xz* and *yz* mirror planes. We assign the dipolar symmetry in Fig. 2a to transient transverse lattice displacement along the *z*-direction and the quadrupolar symmetry in Fig. 2b to longitudinal lattice displacement along the in-plane *x-y* direction, as detailed below.

The derivation of spin generating torque from lattice displacement has been reported by Shen and Bauer [20]. Briefly, laser excitation can transiently heat the sample and induce lattice displacement. Through magnetoelastic coupling (constant *b*), lattice displacement can impose shear and pressure stress to magnetizations and the direction of the induced torque depends on the direction of lattice displacement. The torque or effective field $H_T$ resulting from longitudinal ($R_l$), transverse ($R_t$), and out-of-plane displacements ($R_z$) can be written as [20]:

$$H_T = i\left(\frac{bk}{\hbar\gamma\mu_0}\right)\left[\hat{x}\left(R_l \sin(2\theta) + R_t \cos(2\theta)\right) + \hat{z}R_z \cos(\theta)\right] \quad (1).$$

$k$: in-plane wavevector, $b$: magnetoelastic coupling constant, $\gamma$: gyromagnetic ratio, $\theta$: in-plane angle between the equilibrium magnetic moment and $k$, $\mu_0$: Bohr magneton, $\hbar$: Planck constant, $\hat{x}$ and $\hat{z}$: unit vectors in $x$ and $z$ directions. The light propagation and the external magnetic field $B_0$ are both along the $z$-axis. Using parameters for CrSBr, we simulate the magnetization maps resulting from magnetoelastic coupling in the presence of shear stress due to an in-plane longitudinal phonon mode and pressure stress from an out-of-plane transverse phonon mode, respectively. The simulated patterns induced by the shear and pressure stress, Fig. 2c and Fig. 2d, show the same symmetries as the experimental data at 20 ps and 100 ps, respectively. As a result, we assign the distinct symmetries in magnetization maps as arising from in-plane longitudinal and out-of-plane transverse lattice displacements, respectively (Fig. S1). Note that, in addition to the dipolar and quadrupolar patterns, the simulation also reveals circular patterns at larger distances with the same symm https://ocha.facilities.columbia.edu/subleases etries. These circular patterns at longer distances are not observed in Fig. 2a and 2b, likely resulting in part from insufficient experimental sensitivity and in part from incoherent lattice displacements from multiple phonon modes involved.

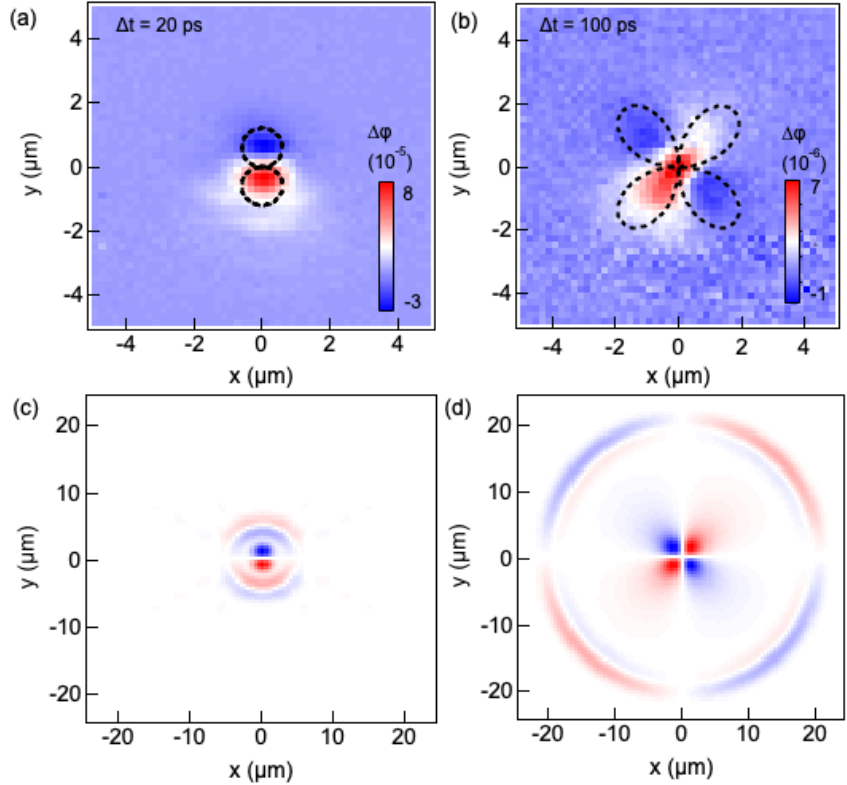

**Fig. 2. Transient magnetization map reflecting magnetoelastic coupling.** Experimentally measured magnetization maps from tr-MOKE at pump-probe delays (a) $\Delta t = 20$ ps and (b) $\Delta t = 100$ ps. Calculated magnetization map from (c) shear and (d) pressure stress. See details in SI.

We find from time-domain measurements that the fast decay in shear stress due to an in-plane longitudinal phonon mode may contribute to an additional channel for the delayed launch of coherent magnons. Fig. 3a shows time-dependent MOKE signal obtained from spatially overlapping pump-probe pulses. The presence of coherent magnons is obvious in the periodic oscillations of the magnetization. We carry out Fourier transform (FT) directly from the time domain signal, Fig. 3b, which reveals two magnon modes at 24.6 ± 0.8 GHz (time period $\tau_1$ = 42 ps) and 33.4 ± 0.5 GHz (time period $\tau_2$ = 30 ps), assigned to the optical (Op) and acoustic (Ac) modes of the interlayer AFM magnon, respectively. This assignment originates from the specific

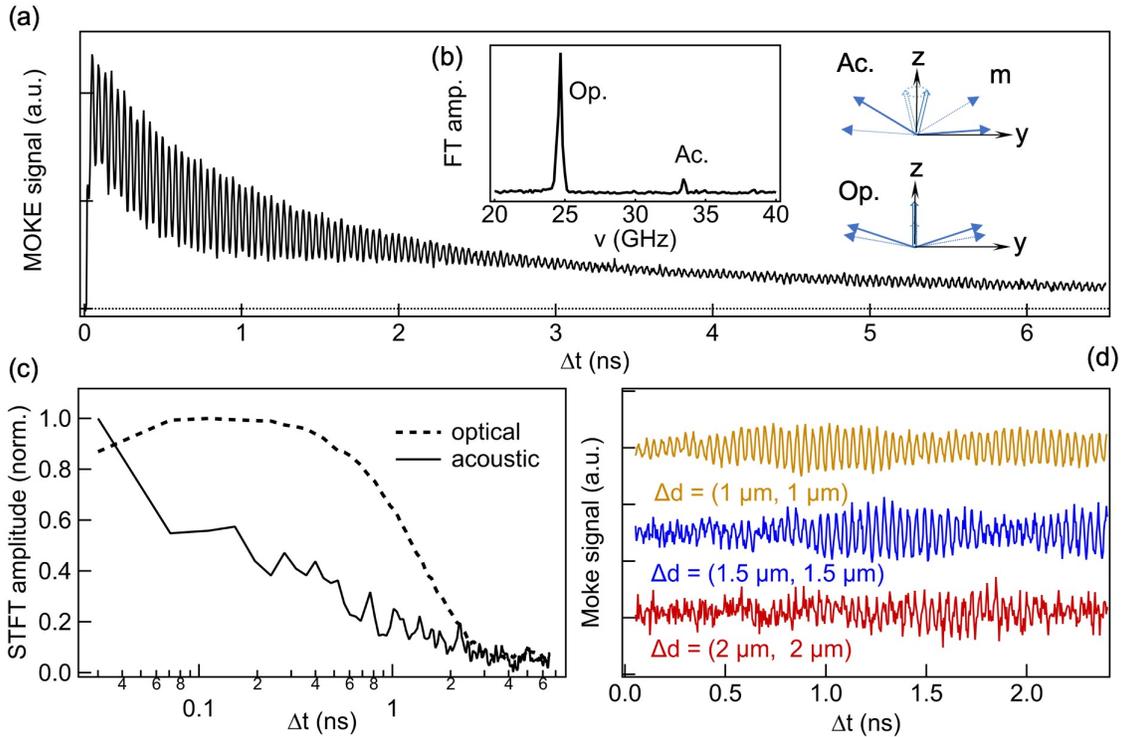

**Fig. 3. Temporal dynamics of optical and acoustic magnon modes.** (a) the kinetic trace of tr-MOKE signal with the inset showing early time dynamics. (b) normal modes of acoustic and optical magnon modes(e) short-time Fourier transform (STFT) of (a) obtained with a time window of 50 ps,

anisotropy of the system in the presence of an external magnetic field along the c axis [13,14], as detailed in Fig. S2. The full-width-at-half-maximum (FWHM) of the two modes are 0.8 and 0.5 GHz for optical and acoustic modes, respectively. The narrow linewidth in the frequency domain originates from long-lived coherence in the time-domain signal, indicating low decoherence rates. From antiferromagnetic resonance spectroscopy (AFRS), we find that the intrinsic damping rates of the two modes at momentum vector k ~ 0 are of the order of 0.1 GHz (Figs. S3-S4). However, optically launched coherent magnons possess finite momentum vector distributions as determined by the finite size of the pump laser beam (see below). The momentum vector distribution gives rise to broader width (from decoherence) than that determined in AFRS.

By taking the short-time Fourier transform (STFT, time window 50 ps) of the data in Fig. 3a, we find that the frequencies of the two magnon modes remain constant but their amplitudes (Fig. 3c) show distinctly different time dependences. While the amplitude of the acoustic magnon mode (solid curve) decays monotonically, with a time constant of ~ 0.3 ns, that of the optical magnon mode (dashed curve) shows an additional small rise in < 80 ps and decays on a much longer time scale of ~ 2 ns. The time scale of the additional rise in the optical magnon model (Fig. 3c) is of the same order as the decay time of the in-plane longitudinal phonon mode inferred from the MOKE images in Fig. 2. If this decay occurs on sufficiently short time scales, $t \leq \tau_I$ (= 42 ps), the resulting transient torque may additionally launch a coherent wavepacket of the optical magnons at 24 GHz, thus accounting for the small early-time rise in the amplitude of this mode, Fig. 3c. This time scale is in the range for the relaxation of optical phonons to acoustic phonons [21]. It is likely that the transient torque comes from the conversion of longitudinal optical (LO) phonons to out-of-plane transverse acoustic (TA) phonons.

Supporting the above interpretation, we note the presences of beating patterns in time domain traces. There are minima in the oscillating amplitudes at $\Delta t$ = 3-4 ns for the spatially overlapping pump-probe data in Fig. 3a, and at $\Delta t$ ~ 1.6 ns (brown) and ~ 1.9 ns (blue) for spatially displaced pump-probe at $(\Delta x, \Delta y)$ = (1 μm, 1μm) and (1.5 μm, 1.5 μm), respectively, in Fig. 3d; note that for $(\Delta x, \Delta y)$ = (2 μm, 2μm), the beating pattern is not well resolved due to the low signal-to-noise ratio (red). These beating patterns suggest the presence of closely spaced frequency components. The initial Gaussian spatial profile of the focused excitation pulse creates a demagnetization field whose gradient launches coherent magnon wavepackets in a momentum vector range of approximately 0 - 1/σ, where σ is the Gaussian width (= 1.2 μm in the present case) [22,23]. We estimate from these momentum windows the corresponding frequency range of ~1 GHz based on the dispersions of the optical magnon mode reported by Sun et al. [15]. For the additional delayed launch of the coherent magnon wavepacket, the spatial extent of the torque is expected to be larger than that of the initial laser excitation spot due to the diffusion of carriers and phonons. As a result, we expect the momentum vector range for the additional coherent phonon wavepacket to be narrower than that of the initially launched wavepacket. The difference in the momentum vector (frequency) ranges of the wavepackets provides a plausible explanation for the beating patterns in time domain data.

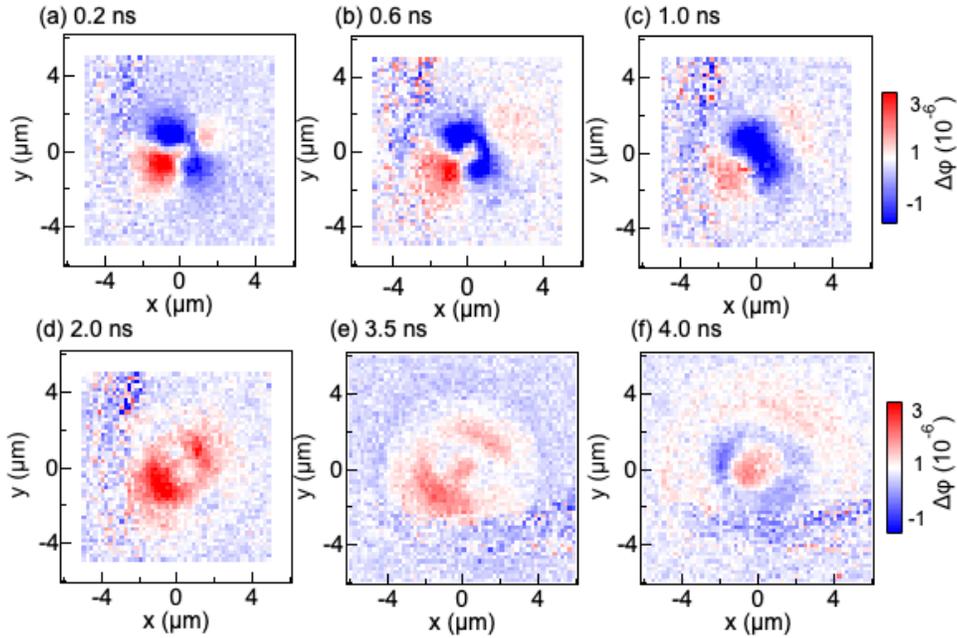

**Fig. 4. Spatial symmetries of magnetization from tr-MOKE images.** The images were obtained at pump-probe delays of Δ$t$ = (a) 0.2 ns, (b) 0.6 ns, (c) 1.0 ns (d) 2.0 ns, (3) 3.5 ns and (f) 4.0 ns following diffraction limited excitation at x, y = 0, 0.

We extend the MOKE imaging in Fig. 2a and 2b to longer time scales, Fig. 4. Contrary to the fast-decaying shear stress responsible for the dipolar magnitization pattern, the pressure stress associated with the quadrupolar pattern persists on a longer time scales, Δt = 0.2, 0.6, 1.0 ns in Fig. 4a, 4b, and 4c, respectively. At longer times, Δt ≥ 2.0 ns, the magnetization pattern becomes increasingly isotropic (circular) (Δ$t$ = 2.0 - 4.0 ns, Fig. 4d-f). We determine group velocities of the two coherent magnon modes along the specific crystal axes by displacing the probe beam with respect to the pump beam and obtain time-domain traces similar to those in Fig. 3d. We then take the STFT with a time window of 50 ps and identify the $\Delta t$ at maximum STFT amplitude. From the analysis, we obtain group velocities from time-resolved MOKE data to be 1.3±0.2 km/s along both a- and b-axes for the optical magnon mode. For the acoustic mode, the group velocity is 3.7±0.4 km/s along the a-axis and not measurable along the b-axis. These results are in good agreement with prior measurements based on exciton sensing [13]. Such isotropic propagation of the optical magnon mode at 24 GHz has been identified by Sun et al. as originating from long-range dipolar interactions in the layered CrSBr [15]. These authors concluded that the dipolar mechanism should be responsible for the fast and isotropic transport of magnons in vdW magnets in general.

## 4. SUMMARY

We identify laser-induced transient strain field in launching coherent magnons, an acoustic and an optical mode, in CrSBr. From spatially and temporally resolved MOKE measurements, we identify two types of transient torque due to magnetoelastic coupling from longitudinal and transverse lattice displacements, leading to distinct magnetization symmetries. The former is short lived (~20 ps) whereas the latter slowly decays over the time scale of nanoseconds. The decay of the short-lived longitudinal phonon may add to an additional but delayed mechanism for the launching of the coherent wavepacket of the optical magnon. These results provide evidence for strong magnetoelastic coupling, which in our measurement is in the non-resonant region, as the acoustic phonon frequencies [24] in the momentum range determined by the spatial extent of the laser excitation spot are an order of magnitude lower than those of the magnons measured here.

Magnetoelastic coupling has been probed extensively before in AFM materials [25], such as the three-dimensional NiO [26] and α-MnTe [27]. Conventionally, magnetoelastic coupling stems from modulation in magnetic anisotropy or spin-orbit coupling (SOC) interactions by collective lattice oscillations. In CrSBr, magnetic anisotropy stems from anisotropic super-exchange interactions mediated by Br, where the SOC of Br generates a magneto-crystalline anisotropy larger than that of single-ion anisotropy of Cr(III). The large magnetoelastic coupling has been observed in two orther 2D antiferromagnet, $FePS_3$, but it primarily comes from the single-ion anisotropy of Fe(II) [28]. Magnetoelastic coupling has also been reported to be important to a 2D stripe AFM, CrOCl [29]. We posit that their low dimensionality conserves spin interactions to in-plane directions and may play a crucial role in enhancing magnetoelastic coupling. Furthermore, in CrSBr which possesses strong magnetoelectronic and magnetoelastic coupling due to the simultaneous presence of magnetic and semiconducting properties, their common connection to the electronic structure may play a role in enhancing magnetoelastic coupling.

**Acknowledgement**

The spectroscopic and imaging work was supported by the Materials Science and Engineering Research Center (MRSEC) through NSF grant DMR-2011738 to XYZ and XR. The synthesis of the CrSBr crystals was supported as part of Programmable Quantum Materials, an Energy Frontier Research Center funded by the US Department of Energy (DOE), Office of Science, Basic Energy Sciences (BES), under award DE-SC0019443 to XR, XYZ, and XX. The magnetic resonance spectroscopy work was supported by the Air Force Office of Scientific Research under grant FA9550-19-1-0307 to AK. The magnetoelastic coupling calculation is supported by the National Natural Science Foundation of China (Grants No.11974047 and No.12374100) and the Fundamental Research Funds for the Central Universities to KS.

SUPPLEMENTARY INFORMATION

**Transient Magnetoelastic Coupling in CrSBr**

Youn Jue Bae,[1,*,†] Taketo Handa,[1] Yanan Dai,[1] Jue Wang,[1] Huicong Liu,[5] Allen Scheie,[2] Daniel G. Chica,[1] Michael E. Ziebel,[1] Andrew D. Kent,[3] Xiaodong Xu,[4] Ka Shen,[5] Xavier Roy,[1] Xiaoyang Zhu[1,*]

[1] Department of Chemistry, Columbia University, New York, NY 10027, USA
[2] MPA-Q, Los Alamos National Laboratory, Los Alamos, NM 87545, USA
[3] Department of Physics, New York University, New York, NY 10003, USA
[4] Department of Physics, University of Washington, Seattle, WA 98195, USA
[5] The Center of Advanced Quantum Studies and Department of Physics, Beijing Normal University, Beijing 100875, China

**Magnetoelastic coupling induced strain fields.**

The magnetoelastic coupling in antiferroamgnet is expressed by: [1]

$$H_{me} = \sum_{\alpha,\beta} B^{\alpha\beta} n_i^\alpha n_i^\beta \varepsilon^{\alpha\beta}, \quad (1)$$

where $n_i^\alpha$ is α component of the Néel vector, $\varepsilon^{\alpha\beta} = \frac{1}{2}(\frac{\partial R_\alpha}{\partial r_\beta} + \frac{\partial R_\beta}{\partial r_\alpha})$ is the starin tensor, $B^{\alpha\beta} = B^\parallel \delta^{\alpha\beta} + B^\perp(1 - \delta^{\alpha\beta})$ is the magnetoelastic coupling constant. We consider the second-order terms about magnon and phonon in magnetoelastic coupling. The effective field $H_T$ provided by magnetoelastic coupling acting on magnetic moments $m_1$ and $m_2$ is

$$\mu_0 H_{T1}^x = -\mu_0 H_{T2}^x = -B^\perp M_s \varepsilon^{xy}, \quad (2)$$

$$\mu_0 H_{T1}^z = -\mu_0 H_{T2}^z = -B^\perp M_s \varepsilon^{yz}. \quad (3)$$

Following the method from previous work, [2] we can get the transverse displacement $R_z$ and longitudinal displacement $R_l$ induced by laser and calculate the strain tensor $\varepsilon^{xy}$ and $\varepsilon^{yz}$

$$\varepsilon^{xy} = \frac{1}{2} ikR_l \sin 2\theta, \quad \varepsilon^{yz} = \frac{1}{2} ikR_z \cos\theta \quad (4)$$

where θ is the angle between the equilibrium magnetic moment and wave vector **k**. The parameters used in calculation are shown in Table S1. We plot the directions of magnetization, torque, and effective field stemming from transverse and longitudinal lattice displacements. Torque is symmetric (antisymmetric) and antisymmetric with respect to the yz and xz mirror planes from transverse (longitudinal) displacement (Fig S1).

Table S1. Parameters used for this calculation

| Gilbert damping | 0.1 | Anisotropy field along b-axis | 1.6 T |
|---|---|---|---|
| Thickness | 150 nm | Anisotropy field along a-axis | 0.64 T |
| Transverse sound velocity[1] | 1.1 km/s | Magnetization | 2.05 x 10⁵ A/m |
| Longitudinal sound velocity[1] | 5 km/s | Dipolar interaction field | 0.26 T |
| Spot size | 1.4 | External magnetic field (along b-axis) | 0.1 T |
| Interlayer exchange field | 0.2 T | Timer after laser pulse | 4 ns |

[1]Note: These values are from calculated phonon dispersion from 2D materials cloud database. [3]

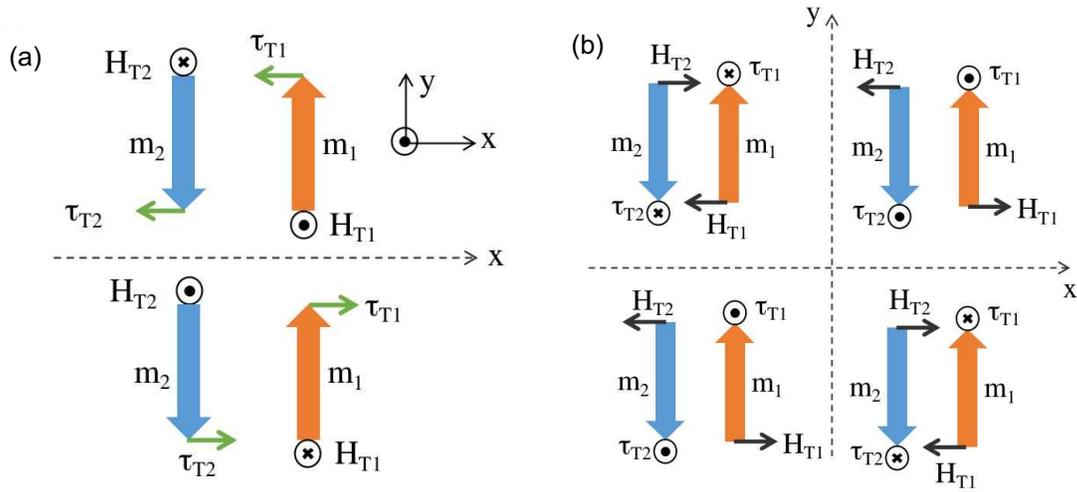

Fig. S1. Directions of magnetization, torque, and effective field stemming from (a) transverse and (b) longitudinal lattice displacements.

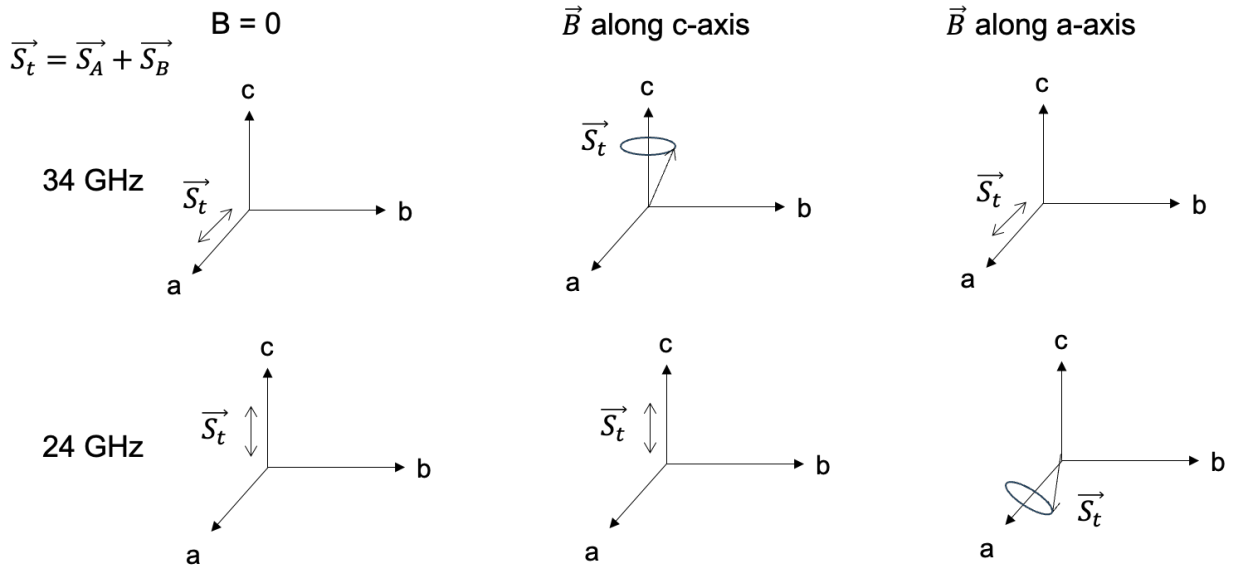

Fig S2. Projected spin rotation and oscillation direction at zero field and external magnetic field along c- and a- axis. Because of the triaxial magnetic anisotropy in CrSBr, at zero field, spin oscillation along a and c axis create two non-degenerate magnon frequency.

**Quantifying magnon damping from antiferromagnetic resonance spectroscopy (AFMR).** We collected magnetic resonance spectra using a broadband co-planar waveguide method in a Quantum Design Physical Property Measurement System (PPMS) at variable temperatures and variable magnetic fields. An exfoliated CrSBr bulk sample (5 mm by 2 mm) is mounted on a co-planar waveguide where an a.c. current provides a small oscillating Oersted field that drives the magnetization into a small-angle precession. At the resonance condition, the amplitude of the precession is maximized. An external d.c. magnetic field is applied perpendicular to the sample plane. The frequency of the a.c. the field is fixed while the d.c. magnetic field is swept from 3 T to zero with a step size of 0.02 T. Both the transmission and reflection spectra are collected using a vector network analyzer. The real part of the transmission port is plotted with respect to the swept magnetic field (Fig S3). A linear combination of symmetric and antisymmetric Lorentzian functions is used to extract the peak position and linewidth of the resonance peaks:

$$\frac{S + A * (H - H_0)}{(H - H_0)^2 + \left(\frac{w}{2}\right)^2} \quad (5)$$

Here S and A are the amplitudes of symmetric and antisymmetric Lorentzian peak, $H$ is the external magnetic field and $H_0$ is the resonance field, and $w$ is the full width half maximum linewidth. The fits are shown in Fig S1. By plotting the frequency vs. $w$ (Fig S4) and fitting the following equation, we extract the coherence time or Gilbert damping [4] of the in-phase mode:

$$\mu_0 \Delta H = \frac{4\pi\alpha f}{\gamma} + \mu_0 \Delta H_0 \quad (6)$$

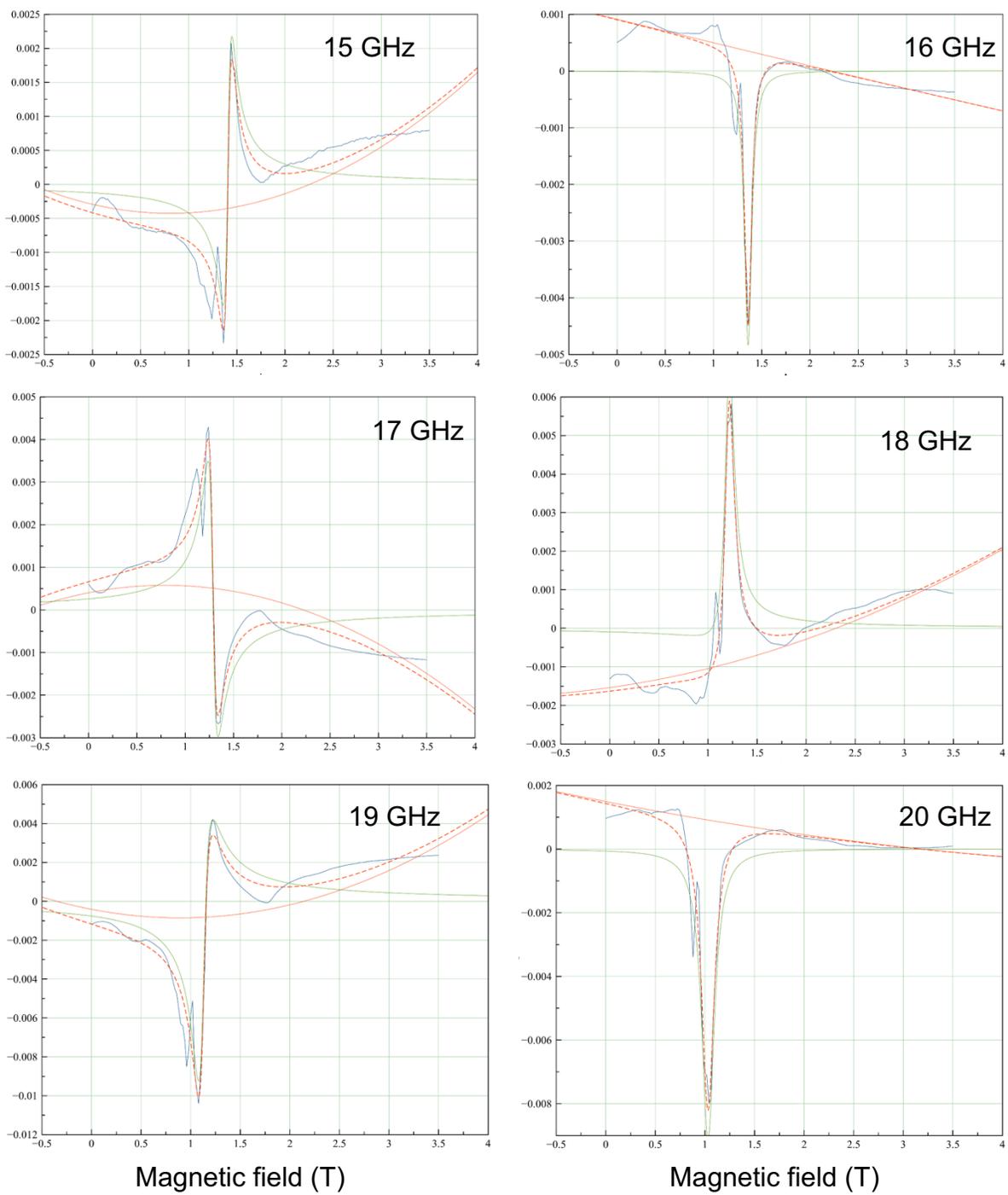

Fig S3. AFMR spectra at different frequencies at 5K.

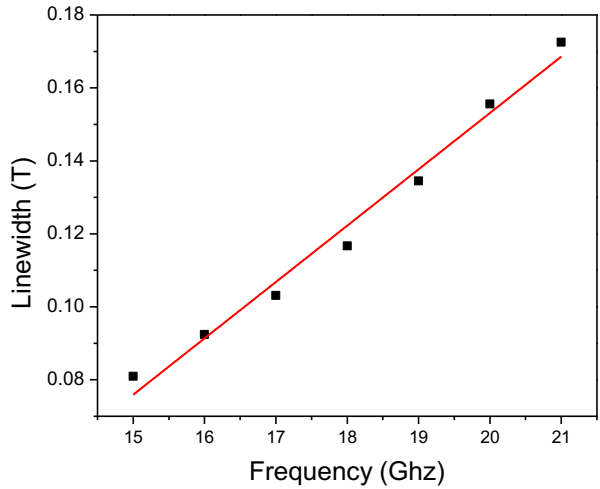

Fig S4. Linewidth vs. frequency plot at 5K. The linewidth at different frequencies is obtained using eq.1 in the main text. Here, the slope is 0.01544 $\pm$ 0.0008 T/GHz.